\documentclass[aps,prl,twocolumn,letterpaper,superscriptaddress,showpacs]{revtex4-2}
\usepackage{amsmath, amsthm}
\usepackage{keyval}
\usepackage{graphicx,latexsym}
\usepackage{dcolumn}
\usepackage{bm}
\usepackage{color}
\usepackage{url}
\usepackage{CJK}
\usepackage{float}
\usepackage{hyperref}
\usepackage{setspace}

\begin{document}
\title{Pairing Properties of the $t$-$t'$-$t''$-$J$ model}

\begin{CJK*}{UTF8}{}
\author{Shengtao Jiang(\CJKfamily{gbsn}蒋晟韬)}
\altaffiliation{shengtaj@uci.edu}
\affiliation{Department of Physics and Astronomy, University of California, Irvine, California 92697, USA}

\author{Douglas J. Scalapino}
\altaffiliation{djs@physics.ucsb.edu}
\affiliation{Department of Physics, University of California, Santa Barbara, California 93106, USA}

\author{Steven R. White}
\altaffiliation{srwhite@uci.edu}
\affiliation{Department of Physics and Astronomy, University of California, Irvine, California 92697, USA}
\date{\today}

\begin{abstract}
We study the pairing properties of the two-dimensional $t$-$t'$-$t''$-$J$ model, where $t'$ and $t''$ are second and third neighbor hoppings, at a doping level $x\approx0.1$. Recent studies of the $t$-$t'$-$J$ model find strong pairing for $t'>0$, associated with electron doping, but an absence of pairing for $t'<0$ associated with hole doping. This is in contrast to the cuprates, where the highest transition temperatures appear for hole doping. Model parameterizations for the cuprates estimate a  $t''$  comparable to $t'$, which, in principle, might fix this discrepancy. However, we find that it does not; we observe a suppression of pairing for the hole-doped system ($t'<0, t''>0$) while for the electron-doped system ($t'>0, t''<0$) $d$-wave pairing is robust.  Extended hoppings appear to be insufficient to make the one-band $t$-$t'$-$J$ model capable of describing the pairing in the hole doped system. 
\end{abstract}
\maketitle
\end{CJK*}

Can a single-band two dimensional $t$-$t'$-$t''$-$J$ model capture the physics of both  the hole and electron doped high-$T_c$ cuprates?  This model, with near-neighbor $t$, next nearest neighbor $t'$ and third nearest neighbor $t''$ hopping parameters, a near neighbor spin exchange $J$ and the restriction of no double-site occupancy, has been found to describe a number of properties seen in these materials. For example: 
(1) the asymmetrical behavior of the commensurate antiferromagnetic(AFM) spin correlations which rapidly decrease with hole doping ($t'<0,t''>0$) but remain strong for a much longer range of electron doping ($t'>0,t''<0$)\cite{tohyama1994,tohyama2004}; 
(2) the appearance of stripes in the hole doped region and their absence in the low electron-doped region\cite{white1999,scalapino2012}; and 
(3) a single particle spectral weight where doped holes lead to the appearance of Fermi arcs around the ($\pm \pi/2, \pm \pi/2$) regions of the Brillouin zone while doped electrons are accommodated near the ($\pm \pi,0$) and ($0,\pm \pi$) regions\cite{tohyama1994,tohyama2004}.

However, while the material itself has clear superconductivity with a high transition temperature, it has been unclear whether these models have a superconducting ground state. In many cases the $t$-$J$ model and its ``parental'' Hubbard model have competing/intertwined orders\cite{vmc2021,Corboztj2014,corboz2011stripes,4leghub-yfjiang,4leghub-huang,4legtj-dodaro,4legtj-hcjiang,stripe-vmc1,stripe-vmc2,stripe-afqmc,vmctjstripe,hubstp-Ido,hubstp-tocchio,senechal2005,machida1,xu2020competing,himeda2002stripe,jiang2022stripe,xu2022stripes,huang2022intertwined,mai2022intertwined,hub-bench2015,stripehubbard} particularly stripes and uniform $d$-wave superconductivity, that are very close in energy. 
Recently, density-matrix renormalization group (DMRG) studies found that while $t'>0$ (electron doping) can lead to strong and unambiguous $d$-wave superconductivity in the ground state\cite{tt'j,sheng,jiang2021high}, for $t'<0$ (hole doping) the model exhibits charge stripes with the pairing suppressed\cite{tt'j}. 
While the coexistence of superconductivity and short range AFM correlations have been reported on the electron doped side\cite{greene2020strange},  the  absence  of  a  superconducting phase  for  the  hole  doped  case, is clearly at odds with the long held belief that the $t$-$t'$-$J$ model provides an appropriate  model for understanding the cuprate superconductors.

The third neighbor hopping $t''$ is thought to be roughly the same size as $t'$ in the cuprates \cite{lee2006doping,tohyama2004,leung1997,xiang1996}. For the case of the hole-doped cuprates, it reflects the extended nature of the Zhang-Rice singlet\cite{zhangrice} of the 3-band CuO$_2$ model\cite{sasha-tt'J}. Can the addition of $t''$ fix the discrepancy?
In this Letter, we use DMRG to investigate pairing properties of the $t$-$t'$-$t''$-$J$ model at a doping level $x\approx0.1$.  Our main conclusion is that $t''$ does not resolve the discrepancy. We find that the parameters used for the electron-doped cuprates ($t'>0, t''<0$) enhance superconductivity, both individually and in combination. However the ones used for the hole-doped cuprates ($t'<0, t''>0$) suppress it. In most of the region with pairing there is coexisting antiferromagnetic(AFM) order and uniform electron/hole density, i.e. an absence of charge stripes.
These results imply that the extended $t$-$t'$-$t''$-$J$ model fails to capture the superconducting phase of the hole-doped cuprates. 

\begin{center}
\begin{figure*}[ht]
	\vspace{-0.0cm}
    \includegraphics[width=1.7\columnwidth,clip=true]{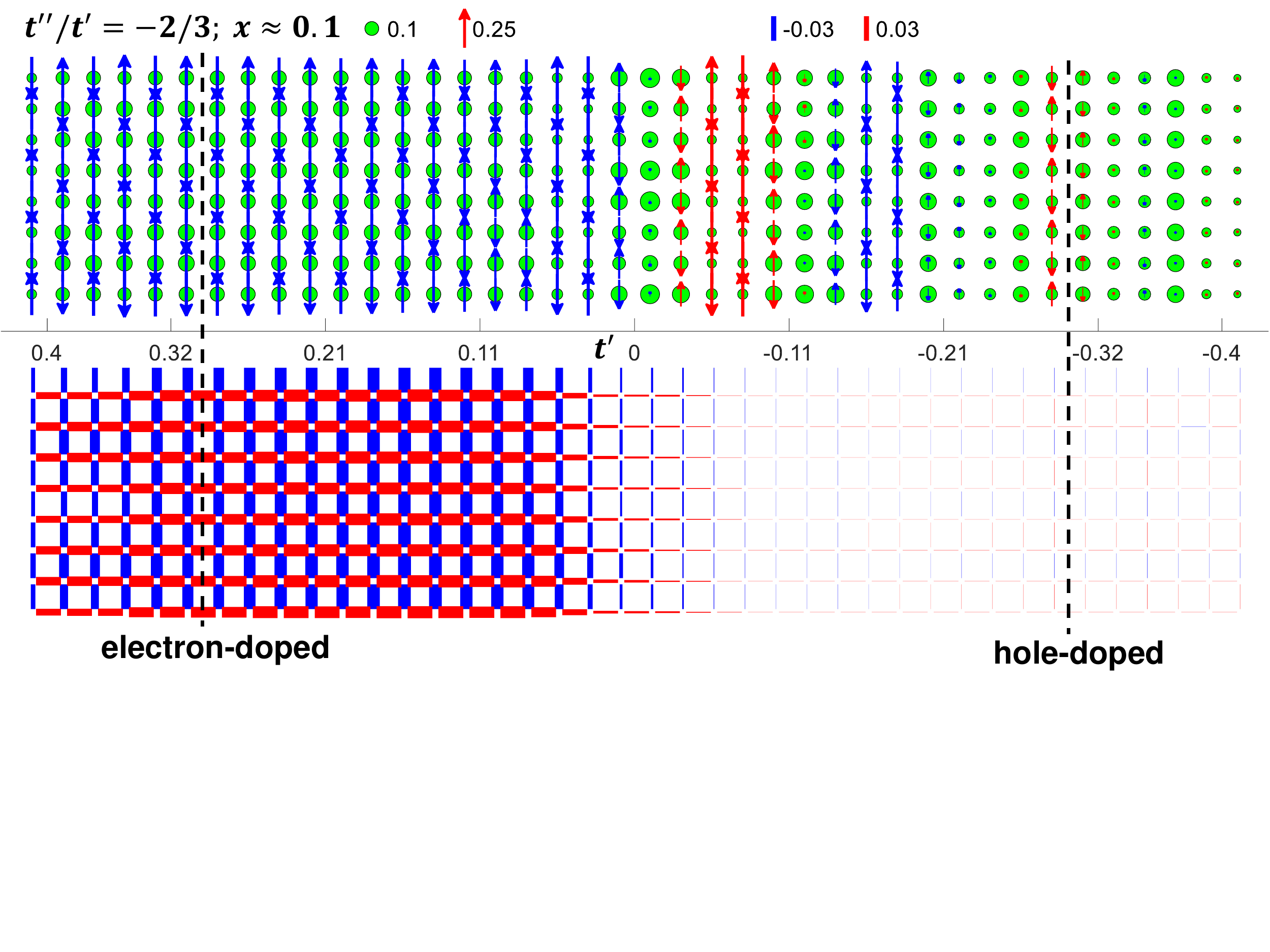}
	\vspace{-3.3cm}
	\caption{
		\label{fig:scan1}
         A scan where both $t''$ and $t'$ vary with a fixed ratio of $-2/3$. The system crosses from a parameter region corresponding to electron-doping to one corresponding to  hole-doping\cite{lee2006doping,leung1997}. In the upper plot, the area of the circles represents the local fermion density $x_i$ at site $i$ such that for the electron doped system, the electron density $n_i=1+x_i$ and for the hole doped system $n_i=1-x_i$, the arrows represents the local $\langle S^z \rangle $ with the colors indicating different AFM regions. In the lower plot the width/color of the links represent the magnitude/sign of the singlet pairing between two adjacent sites. The electron-doped region has coexisting AFM, uniform hole density and strong $d$-wave pairing while the hole doped region is striped with the pairing suppressed.}
	\vspace{-0.2cm}
\end{figure*}
\end{center}

\vspace{-1.0cm}
The Hamiltonian of the $t$-$t'$-$t''$-$J$ model that we will study is:
\begin{equation}
\begin{aligned}
    \label{eq:tj}
    H=&\sum_{\langle ij \rangle \sigma}-t(c^\dag_{i\sigma}c_{j\sigma}+h.c.)\\
    +&\sum_{\langle \langle ij \rangle \rangle\sigma}-t'(c^\dag_{i\sigma}c_{j\sigma}+h.c.)
    +\sum_{\langle\langle\langle ij \rangle \rangle\rangle\sigma}-t''(c^\dag_{i\sigma}c_{j\sigma}+h.c.)\\
    +&\sum_{\langle ij \rangle}J(\Vec{S_i}\cdot \Vec{S_j}-\frac{1}{4}n_i^{tot} n_j^{tot})+\sum_i -\mu n_i^{tot}
\end{aligned}
\end{equation}
with the restriction of no double site occupancy.
The single/double/triple brackets under the summations denote first/second/third nearest neighbor pairs of sites, and $n^{tot}_i=n_{i\uparrow}+n_{i\downarrow}$ is the total particle density on site $i$. The nearest neighbor hopping $t$ is set to 1 and the spin exchange $J$ is set to 0.4 for all calculations.  It has been proposed\cite{lee2006doping,leung1997} that the parameters $t'=-0.3$, $t''=0.2$ correspond to hole-doped cuprates. Under a particle-hole transformation which changes the signs of both hoppings, the set ($t'=0.3, t''=-0.2$) corresponds to electron doping with a filling $n=1+x$. Here we treat both $t'$ and $t''$ as adjustable parameters to investigate their individual and combined effects on superconductivity. A chemical potential $\mu$ is used to control the doping level $x$. We keep $x\approx 0.1$ where with $t''=0$ it is known to have one phase for $t'>0$(electron doping) with clear superconductivity and another phase for $t'<0$(hole doping) where superconductivity is suppressed\cite{tt'j}. 

The DMRG calculations are carried out using the ITensor library\cite{itensor}. We typically perform around 20 sweeps and keep a maximum bond dimension $m\sim 3000$ which provides good convergence for local observables of interest: the local doping on site $i$: $n_{dope}(i)=1-n^{tot}_i$, the local magnetization on site $i$: $S^z(i)=\frac{1}{2}(n_{i\uparrow}-n_{i\downarrow})$ and the local singlet pairing order parameter $\Delta^(i,j)$ on nearest neighboring sites $i$ and $j$: $ \Delta^\dag(i,j)=\frac{1}{\sqrt{2}}(c^\dag_{i,\uparrow}c^\dag_{j,\downarrow} - c^\dag_{i,\downarrow}c^\dag_{j,\uparrow})$. Note that this bond dimension would not be sufficient for more difficult observables, such as long-range pairing correlations\cite{sheng,jiang2022stripe}.  Our main interest is the overall phase diagram. We allow symmetry to break in these phases (encouraging the breaking with initial states or boundary fields) so that local observables describe them. Although our results are limited to width-8 cylinders, there are no indications that larger widths would give a qualitatively different phase diagram.

\begin{center}
\begin{figure*}[t]
	\vspace{-0.0cm}
    \includegraphics[width=1.7\columnwidth,clip=true]{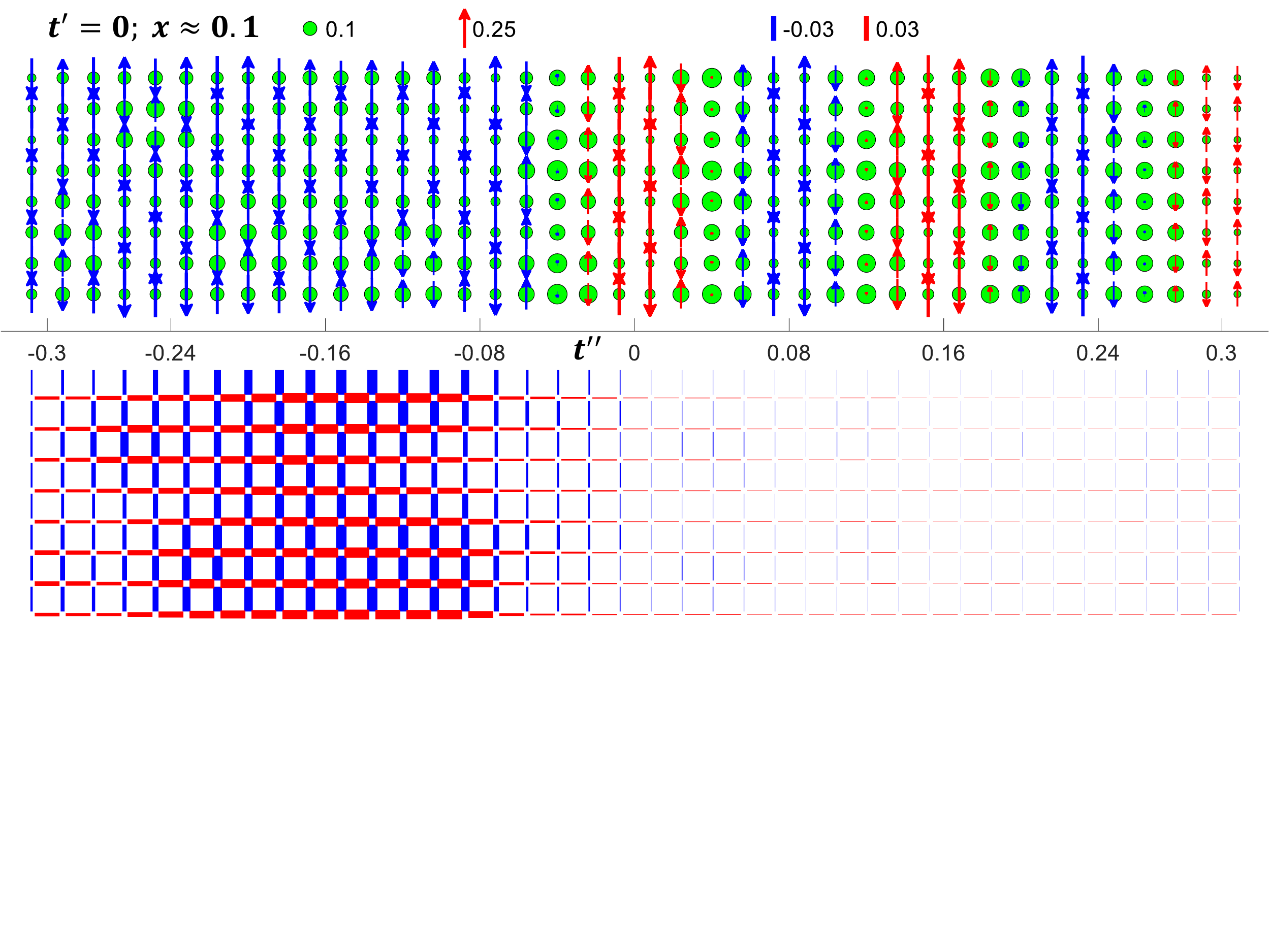}
	\vspace{-4.0cm}
	\caption{
		\label{fig:scan2}
         A $t''$-varying scan at $t'=0$ and $x\approx0.1$ plotted in the same way as in Fig.~\ref{fig:scan1}. For $t''<0$ the system shows AFM order with a slightly nonuniform hole density and $d$-wave pairing that peaks for $t''=-0.16$. For $t''>0$ stripes appear and the pairing is suppressed.}
	\vspace{-0.2cm}
\end{figure*}
\end{center}

\vspace{-1.0cm}
A potential difficulty for DMRG simulations on large systems is the
tendency to get stuck in metastable phases. This issue is most serious near phase boundaries where different competing phases are close in energy. To reduce this problem, we employ ``scan'' calculations where along the length of a long cylinder we vary one or two parameter(s)\cite{tt'j} linearly and slowly. A varying chemical potential is also used to keep the doping level constant---the variation of $\mu$ down the cylinder is manually adjusted with repeated runs to keep the doping constant.   In this way the cylinder forms different phases in different regions along the length of the cylinder. The competition between phases occurs over a few columns of sites which automatically ``regularizes'' or controls the competition so that no large length scales or large entanglement can occur. The converged position of the phase boundary translates to the parameter values of the boundary. 
The phases are particularly stable near the center of each region.  Near the centers of each region, we have also performed separate non-scan/fixed-parameter simulations to verify the phase. 

In Fig.~\ref{fig:scan1} we show a scan calculation which crosses a parameter region in which the system changes from electron-doped to hole-doped cuprates. The upper panel shows the local fermion density $x_i$ and spin $\langle S^z \rangle$ observables which gives an immediate view of the phases and their spin/charge orders.
Restricted by the $t$-$J$ space, we simulate the electron-doped system by staying below half-filling and performing a particle-hole transformation which changes the sign for both $t'$ and $t''$. Therefore, one can treat the fermion density $x_i$(green circles) in the left half of the system($t'>0$) as the density of doped electrons added to the half-filled band and in the right half of the system($t'<0$) as doped holes.
For the parameter set ($t'=0.3,t''=-0.2$) corresponding to the electron-doped case, the system exhibits commensurate AFM with uniform density and a strong $d$-wave pairing shown in lower plot.
The pairing reaches its maximum around $t'\approx 0.2, t''\approx-0.13$ instead of increasing monotonically with $|t'|,|t''|$.
On the other hand, the parameter set ($t'=-0.3,t''=0.2$) associated with the hole-doped system shows a charge density wave with the pairing suppressed. One can see that its spin pattern which has much smaller spin moments differs from the conventional striped state near the $t'=0$ region. As discussed in the supplemental materials\cite{sm}, the spin correlations in this unconventional striped phase are short-ranged, mimicking the behavior of the two-leg Heisenberg ladders.

Next in Fig~\ref{fig:scan2} we present another scan calculation which varies $t''$ with $t'$ fixed at zero to investigate the individual effect of $t''$.  For $t''<0$ corresponding to the electron-doped cuprates we see commensurate AFM with a slight non-uniformity in density, while for $t''>0$ related to hole-doped cuprates it shows conventional stripes. The $d$-wave pairing only exists for $t''<0$ and its magnitude peaks around $t''=-0.16$.  Roughly speaking, the effect of having a $t''$ alone is very similar to having a $t'$ alone\cite{tt'j} with opposite sign. The small differences are that the enhancement of pairing by $t''>0$ is overall weaker and happens over a narrower window. 
The rest of the scan calculations are presented in the supplemental material\cite{sm}.

\begin{center}
\begin{figure}[t]
	\vspace{-0.0cm}
    \includegraphics[width=1.0\columnwidth,clip=true]{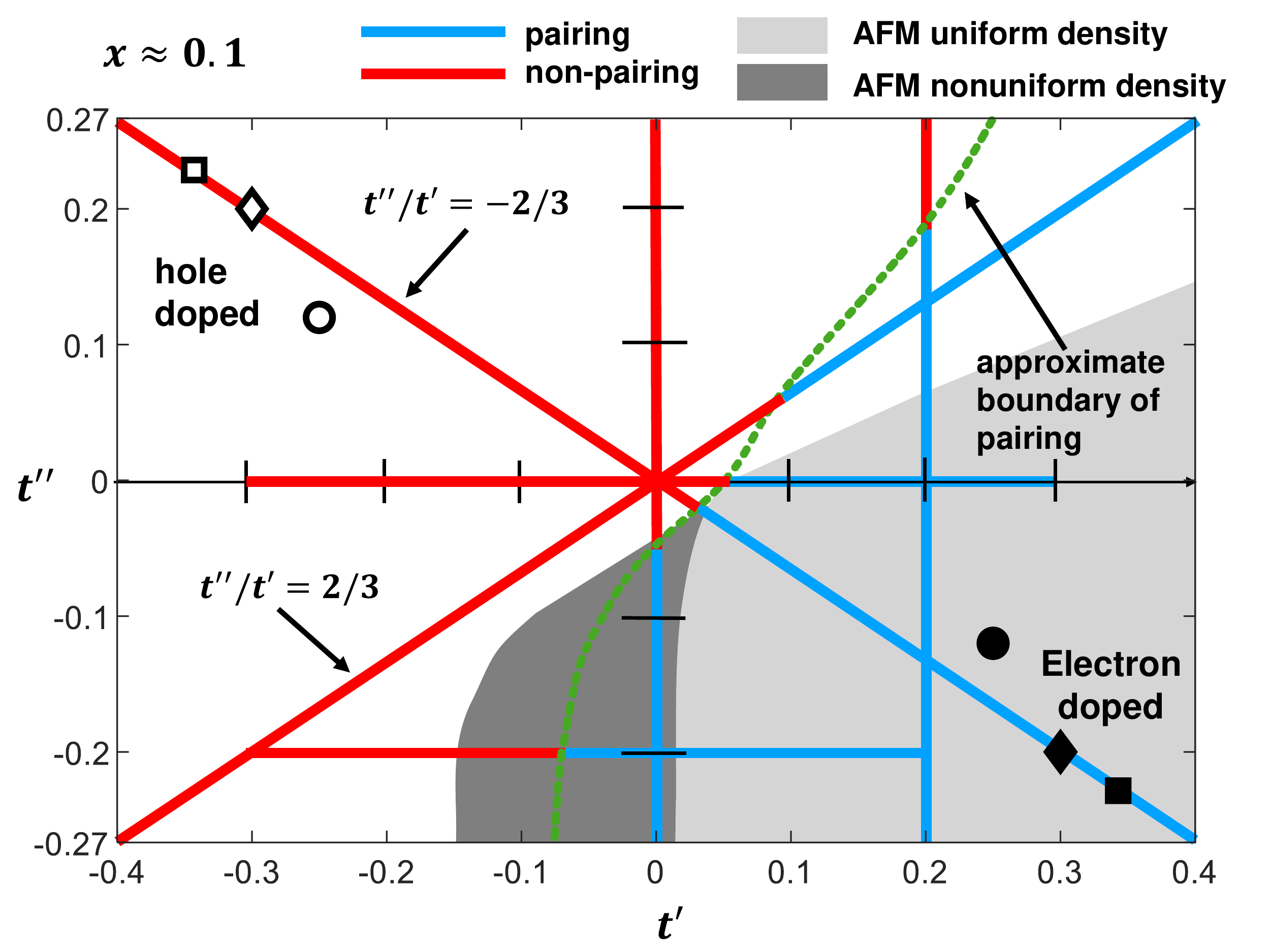}
	\vspace{-0.6cm}
	\caption{
		\label{fig:phasediag}
          An approximate phase diagram in the $t'-t''$ plane at doping $x\approx 0.1$. The lines are ``scans'' with the blue/red color denoting the parameter range with/without  pairing. The dotted green line shows the extrapolated pairing phase boundary based on the scans. The light/dark gray regions have AFM order with uniform/nonuniform hole density while the  white background is striped. Square\cite{xiang1996}, diamond\cite{leung1997,lee2006doping} and circle\cite{tohyama2004} markers indicate the ($t',t''$) values proposed in several studies with solid markers for electron doped systems and hollow markers for hole doped systems.}
	\vspace{-0.2cm}
\end{figure}
\end{center}

By collecting all the scans, we have constructed the approximate phase diagram in the $t'-t''$ plane shown in Fig.~\ref{fig:phasediag}. The solid lines are DMRG scans as discussed above.  The blue parts of these scans denotes parameter ranges with $d$-wave  pairing and the red parts without pairing. By connecting the transition points on these scans, we have mapped out an approximate boundary of the pairing phase, as indicated by the green dotted line. 

It is clear from Fig.~\ref{fig:phasediag} that $t'>0$ and $t''<0$ enhance pairing, while $t'<0$ and $t''>0$ suppress it.  This is true when $t'$\cite{tt'j} or $t''$ acts individually, as can be seen from the $t'=0$ and $t''=0$ axis. When they act together, the effect appears to be largely additive since the approximate pairing phase boundary is roughly orthogonal to the $t''/t'=-2/3$ line.  In other words, one always has a decrease of pairing strength when starting from a point in the pairing phase(within the range shown in Fig.~\ref{fig:phasediag}) and decreasing $t'$ or increasing $t''$.

The region with superconductivity largely has coexisting AFM order with uniform density, as indicated by the light grey area. This supports the idea that charge stripes (either with or without $\pi$ phase shift of AFM order) compete with pairing and the absence of stripes supports good pair mobility and superconducting phase coherence. 
We also note that the longer range hoppings $t'$ and $t''$ will in principle lead to longer range next-nearest $J'$ and third nearest neighbor $J''$ exchange couplings. For example, if one assumes that the exchange coupling $J=0.4$ arises from a Hubbard model with $t=1$ and $U=10$, then for $t'=0.3$, one would have $J'\approx (t'/t)^2J=0.036$. This longer range exchange coupling has a negligible effect on the AFM order and as we have discussed one will have a phase with coexisting $d$-wave superconducting and AFM order arising from electron doping. However, as Jiang and Kivelson\cite{jiang2021high} have discussed, for a larger value of $J'= J/2$ and corresponding $t'= t/\sqrt{2}$, they find that a $d$-wave superconductor arises from electron doping a spin-liquid state.

\vspace{-1.0cm}
The parameters $t' = 0.3, t'' = -0.2$, associated with an electron-doped system and marked by the open diamond in Fig.~\ref{fig:phasediag}, correspond to a state which is deep inside the pairing phase. Alternatively, the parameters $t' = -0.3, t'' = 0.2$, associated with a hole-doped system and marked by the solid diamond in Fig.~\ref{fig:phasediag}, corresponds to a state which is far away from the pairing phase boundary.
We have tried numerous ways to obtain a phase in the hole doped region which is superconducting, but the only way seems to be to apply an unphysical large pairing field throughout the system, and its removal leads to the disappearance of pairing within a few DMRG sweeps. These results indicate a fundamental flaw in the model for a reliable description of the doping dependence of pairing in the cuprates.

{\it Discussion:} If the $t$-$t'$-$t''$-$J$ model is not sufficient to describe superconductivity in
the cuprates, it is natural to look at less renormalized models. One can think of these models as devised in successive steps starting with all electron calculations, constructing a three band model, reducing it to a single-band Hubbard model, and from there, in the large $U$ limit, to a $t$-$t'$-$t''$-$J$ model. Apparently, somewhere in this process the correct modeling of the pairing for the hole doped system is lost. It may be that the restriction of no double occupancy is too severe and a single band Hubbard model will turn out to be adequate. However, another possibility may stem from the reduction of the three band charge-transfer model to a single-band Mott-Hubbard model, and in particular this reduction on the hole doped side.  
In the case of electron doping, this reduction is fairly simple:  the doped electrons dominantly occupy the coppers sites, and so it is reasonable to integrate out the oxygens.  However, on the hole doped side, the doped holes tend to go  onto both the Cu and its surrounding O($p_x,p_y$) orbitals \cite{white2015doping}, and conceptually the reduction to a one band model proceeds through the Zhang-Rice singlet picture\cite{zhangrice}. Although there is evidence from simulation\cite{cui2020,mai2021,arrigoni2009phase}
and experiments\cite{brookes2001detection,tjeng1997spin,harada2002unique} that support the Zhang-Rice singlet picture, it may also be that a more careful treatment of the reduction could introduce new terms beyond extended hoppings, which are necessary to get the pairing behavior right. It may also be that the required terms are complicated or very extended, and it would be better to go back to a three band model.

We thanks R. L. Greene and S. A. Kivelson for useful discussions.
SJ and SRW are supported by the NSF under DMR-2110041.
DJS was supported by the Scientific Discovery through Advanced Computing (SciDAC) program funded by the U.S. Department of Energy. 

\bibliography{ref}

\begin{thebibliography}{45}%
\makeatletter
\providecommand \@ifxundefined [1]{%
 \@ifx{#1\undefined}
}%
\providecommand \@ifnum [1]{%
 \ifnum #1\expandafter \@firstoftwo
 \else \expandafter \@secondoftwo
 \fi
}%
\providecommand \@ifx [1]{%
 \ifx #1\expandafter \@firstoftwo
 \else \expandafter \@secondoftwo
 \fi
}%
\providecommand \natexlab [1]{#1}%
\providecommand \enquote  [1]{``#1''}%
\providecommand \bibnamefont  [1]{#1}%
\providecommand \bibfnamefont [1]{#1}%
\providecommand \citenamefont [1]{#1}%
\providecommand \href@noop [0]{\@secondoftwo}%
\providecommand \href [0]{\begingroup \@sanitize@url \@href}%
\providecommand \@href[1]{\@@startlink{#1}\@@href}%
\providecommand \@@href[1]{\endgroup#1\@@endlink}%
\providecommand \@sanitize@url [0]{\catcode `\\12\catcode `\$12\catcode
  `\&12\catcode `\#12\catcode `\^12\catcode `\_12\catcode `\%12\relax}%
\providecommand \@@startlink[1]{}%
\providecommand \@@endlink[0]{}%
\providecommand \url  [0]{\begingroup\@sanitize@url \@url }%
\providecommand \@url [1]{\endgroup\@href {#1}{\urlprefix }}%
\providecommand \urlprefix  [0]{URL }%
\providecommand \Eprint [0]{\href }%
\providecommand \doibase [0]{https://doi.org/}%
\providecommand \selectlanguage [0]{\@gobble}%
\providecommand \bibinfo  [0]{\@secondoftwo}%
\providecommand \bibfield  [0]{\@secondoftwo}%
\providecommand \translation [1]{[#1]}%
\providecommand \BibitemOpen [0]{}%
\providecommand \bibitemStop [0]{}%
\providecommand \bibitemNoStop [0]{.\EOS\space}%
\providecommand \EOS [0]{\spacefactor3000\relax}%
\providecommand \BibitemShut  [1]{\csname bibitem#1\endcsname}%
\let\auto@bib@innerbib\@empty
\bibitem [{\citenamefont {Tohyama}\ and\ \citenamefont
  {Maekawa}(1994)}]{tohyama1994}%
  \BibitemOpen
  \bibfield  {author} {\bibinfo {author} {\bibfnamefont {T.}~\bibnamefont
  {Tohyama}}\ and\ \bibinfo {author} {\bibfnamefont {S.}~\bibnamefont
  {Maekawa}},\ }\bibfield  {title} {\bibinfo {title} {Role of
  next-nearest-neighbor hopping in the t-t'-j model},\ }\href@noop {}
  {\bibfield  {journal} {\bibinfo  {journal} {Physical Review B}\ }\textbf
  {\bibinfo {volume} {49}},\ \bibinfo {pages} {3596} (\bibinfo {year}
  {1994})}\BibitemShut {NoStop}%
\bibitem [{\citenamefont {Tohyama}(2004)}]{tohyama2004}%
  \BibitemOpen
  \bibfield  {author} {\bibinfo {author} {\bibfnamefont {T.}~\bibnamefont
  {Tohyama}},\ }\bibfield  {title} {\bibinfo {title} {Asymmetry of the
  electronic states in hole-and electron-doped cuprates: exact diagonalization
  study of the t- t'- t''- j model},\ }\href@noop {} {\bibfield  {journal}
  {\bibinfo  {journal} {Physical Review B}\ }\textbf {\bibinfo {volume} {70}},\
  \bibinfo {pages} {174517} (\bibinfo {year} {2004})}\BibitemShut {NoStop}%
\bibitem [{\citenamefont {White}\ and\ \citenamefont
  {Scalapino}(1999)}]{white1999}%
  \BibitemOpen
  \bibfield  {author} {\bibinfo {author} {\bibfnamefont {S.~R.}\ \bibnamefont
  {White}}\ and\ \bibinfo {author} {\bibfnamefont {D.}~\bibnamefont
  {Scalapino}},\ }\bibfield  {title} {\bibinfo {title} {Competition between
  stripes and pairing in a t- t'- j model},\ }\href@noop {} {\bibfield
  {journal} {\bibinfo  {journal} {Physical Review B}\ }\textbf {\bibinfo
  {volume} {60}},\ \bibinfo {pages} {R753} (\bibinfo {year}
  {1999})}\BibitemShut {NoStop}%
\bibitem [{\citenamefont {Scalapino}\ and\ \citenamefont
  {White}(2012)}]{scalapino2012}%
  \BibitemOpen
  \bibfield  {author} {\bibinfo {author} {\bibfnamefont {D.}~\bibnamefont
  {Scalapino}}\ and\ \bibinfo {author} {\bibfnamefont {S.}~\bibnamefont
  {White}},\ }\bibfield  {title} {\bibinfo {title} {Stripe structures in the
  t--t'-j model},\ }\href@noop {} {\bibfield  {journal} {\bibinfo  {journal}
  {Physica C: Superconductivity}\ }\textbf {\bibinfo {volume} {481}},\ \bibinfo
  {pages} {146} (\bibinfo {year} {2012})}\BibitemShut {NoStop}%
\bibitem [{\citenamefont {Marino}\ \emph {et~al.}(2021)\citenamefont {Marino},
  \citenamefont {Becca},\ and\ \citenamefont {Tocchio}}]{vmc2021}%
  \BibitemOpen
  \bibfield  {author} {\bibinfo {author} {\bibfnamefont {V.}~\bibnamefont
  {Marino}}, \bibinfo {author} {\bibfnamefont {F.}~\bibnamefont {Becca}},\ and\
  \bibinfo {author} {\bibfnamefont {L.~F.}\ \bibnamefont {Tocchio}},\ }\href
  {https://doi.org/10.48550/ARXIV.2111.04623} {\bibinfo {title} {Stripes in the
  extended $t-t^\prime$ hubbard model: A variational monte carlo analysis}}
  (\bibinfo {year} {2021})\BibitemShut {NoStop}%
\bibitem [{\citenamefont {Corboz}\ \emph {et~al.}(2014)\citenamefont {Corboz},
  \citenamefont {Rice},\ and\ \citenamefont {Troyer}}]{Corboztj2014}%
  \BibitemOpen
  \bibfield  {author} {\bibinfo {author} {\bibfnamefont {P.}~\bibnamefont
  {Corboz}}, \bibinfo {author} {\bibfnamefont {T.~M.}\ \bibnamefont {Rice}},\
  and\ \bibinfo {author} {\bibfnamefont {M.}~\bibnamefont {Troyer}},\
  }\bibfield  {title} {\bibinfo {title} {Competing states in the
  $\mathit{t}\ensuremath{-}\mathit{J}$ model: Uniform $d$-wave state versus
  stripe state},\ }\href {https://doi.org/10.1103/PhysRevLett.113.046402}
  {\bibfield  {journal} {\bibinfo  {journal} {Phys. Rev. Lett.}\ }\textbf
  {\bibinfo {volume} {113}},\ \bibinfo {pages} {046402} (\bibinfo {year}
  {2014})}\BibitemShut {NoStop}%
\bibitem [{\citenamefont {Corboz}\ \emph {et~al.}(2011)\citenamefont {Corboz},
  \citenamefont {White}, \citenamefont {Vidal},\ and\ \citenamefont
  {Troyer}}]{corboz2011stripes}%
  \BibitemOpen
  \bibfield  {author} {\bibinfo {author} {\bibfnamefont {P.}~\bibnamefont
  {Corboz}}, \bibinfo {author} {\bibfnamefont {S.~R.}\ \bibnamefont {White}},
  \bibinfo {author} {\bibfnamefont {G.}~\bibnamefont {Vidal}},\ and\ \bibinfo
  {author} {\bibfnamefont {M.}~\bibnamefont {Troyer}},\ }\bibfield  {title}
  {\bibinfo {title} {Stripes in the two-dimensional
  $\mathit{t}\ensuremath{-}\mathit{J}$ model with infinite projected
  entangled-pair states},\ }\href@noop {} {\bibfield  {journal} {\bibinfo
  {journal} {Physical Review B}\ }\textbf {\bibinfo {volume} {84}},\ \bibinfo
  {pages} {041108(R)} (\bibinfo {year} {2011})}\BibitemShut {NoStop}%
\bibitem [{\citenamefont {Jiang}\ \emph {et~al.}(2020)\citenamefont {Jiang},
  \citenamefont {Zaanen}, \citenamefont {Devereaux},\ and\ \citenamefont
  {Jiang}}]{4leghub-yfjiang}%
  \BibitemOpen
  \bibfield  {author} {\bibinfo {author} {\bibfnamefont {Y.-F.}\ \bibnamefont
  {Jiang}}, \bibinfo {author} {\bibfnamefont {J.}~\bibnamefont {Zaanen}},
  \bibinfo {author} {\bibfnamefont {T.~P.}\ \bibnamefont {Devereaux}},\ and\
  \bibinfo {author} {\bibfnamefont {H.-C.}\ \bibnamefont {Jiang}},\ }\bibfield
  {title} {\bibinfo {title} {Ground state phase diagram of the doped {Hubbard}
  model on the four-leg cylinder},\ }\href
  {https://doi.org/10.1103/PhysRevResearch.2.033073} {\bibfield  {journal}
  {\bibinfo  {journal} {Phys. Rev. Research}\ }\textbf {\bibinfo {volume}
  {2}},\ \bibinfo {pages} {033073} (\bibinfo {year} {2020})}\BibitemShut
  {NoStop}%
\bibitem [{\citenamefont {Huang}\ \emph {et~al.}(2018)\citenamefont {Huang},
  \citenamefont {Mendl}, \citenamefont {Jiang}, \citenamefont {Moritz},\ and\
  \citenamefont {Devereaux}}]{4leghub-huang}%
  \BibitemOpen
  \bibfield  {author} {\bibinfo {author} {\bibfnamefont {E.~W.}\ \bibnamefont
  {Huang}}, \bibinfo {author} {\bibfnamefont {C.~B.}\ \bibnamefont {Mendl}},
  \bibinfo {author} {\bibfnamefont {H.-C.}\ \bibnamefont {Jiang}}, \bibinfo
  {author} {\bibfnamefont {B.}~\bibnamefont {Moritz}},\ and\ \bibinfo {author}
  {\bibfnamefont {T.~P.}\ \bibnamefont {Devereaux}},\ }\bibfield  {title}
  {\bibinfo {title} {Stripe order from the perspective of the {Hubbard}
  model},\ }\href@noop {} {\bibfield  {journal} {\bibinfo  {journal} {npj
  Quantum Materials}\ }\textbf {\bibinfo {volume} {3}},\ \bibinfo {pages} {1}
  (\bibinfo {year} {2018})}\BibitemShut {NoStop}%
\bibitem [{\citenamefont {Dodaro}\ \emph {et~al.}(2017)\citenamefont {Dodaro},
  \citenamefont {Jiang},\ and\ \citenamefont {Kivelson}}]{4legtj-dodaro}%
  \BibitemOpen
  \bibfield  {author} {\bibinfo {author} {\bibfnamefont {J.~F.}\ \bibnamefont
  {Dodaro}}, \bibinfo {author} {\bibfnamefont {H.-C.}\ \bibnamefont {Jiang}},\
  and\ \bibinfo {author} {\bibfnamefont {S.~A.}\ \bibnamefont {Kivelson}},\
  }\bibfield  {title} {\bibinfo {title} {Intertwined order in a frustrated
  four-leg $\mathit{t}\ensuremath{-}\mathit{J}$ cylinder},\ }\href
  {https://doi.org/10.1103/PhysRevB.95.155116} {\bibfield  {journal} {\bibinfo
  {journal} {Phys. Rev. B}\ }\textbf {\bibinfo {volume} {95}},\ \bibinfo
  {pages} {155116} (\bibinfo {year} {2017})}\BibitemShut {NoStop}%
\bibitem [{\citenamefont {Jiang}\ \emph {et~al.}(2018)\citenamefont {Jiang},
  \citenamefont {Weng},\ and\ \citenamefont {Kivelson}}]{4legtj-hcjiang}%
  \BibitemOpen
  \bibfield  {author} {\bibinfo {author} {\bibfnamefont {H.-C.}\ \bibnamefont
  {Jiang}}, \bibinfo {author} {\bibfnamefont {Z.-Y.}\ \bibnamefont {Weng}},\
  and\ \bibinfo {author} {\bibfnamefont {S.~A.}\ \bibnamefont {Kivelson}},\
  }\bibfield  {title} {\bibinfo {title} {Superconductivity in the doped
  $\mathit{t}\ensuremath{-}\mathit{J}$ model: Results for four-leg cylinders},\
  }\href {https://doi.org/10.1103/PhysRevB.98.140505} {\bibfield  {journal}
  {\bibinfo  {journal} {Phys. Rev. B}\ }\textbf {\bibinfo {volume} {98}},\
  \bibinfo {pages} {140505(R)} (\bibinfo {year} {2018})}\BibitemShut {NoStop}%
\bibitem [{\citenamefont {Ido}\ \emph {et~al.}(2018{\natexlab{a}})\citenamefont
  {Ido}, \citenamefont {Ohgoe},\ and\ \citenamefont {Imada}}]{stripe-vmc1}%
  \BibitemOpen
  \bibfield  {author} {\bibinfo {author} {\bibfnamefont {K.}~\bibnamefont
  {Ido}}, \bibinfo {author} {\bibfnamefont {T.}~\bibnamefont {Ohgoe}},\ and\
  \bibinfo {author} {\bibfnamefont {M.}~\bibnamefont {Imada}},\ }\bibfield
  {title} {\bibinfo {title} {Competition among various charge-inhomogeneous
  states and $d$-wave superconducting state in {Hubbard} models on square
  lattices},\ }\href {https://doi.org/10.1103/PhysRevB.97.045138} {\bibfield
  {journal} {\bibinfo  {journal} {Phys. Rev. B}\ }\textbf {\bibinfo {volume}
  {97}},\ \bibinfo {pages} {045138} (\bibinfo {year}
  {2018}{\natexlab{a}})}\BibitemShut {NoStop}%
\bibitem [{\citenamefont {Tocchio}\ \emph {et~al.}(2016)\citenamefont
  {Tocchio}, \citenamefont {Becca},\ and\ \citenamefont
  {Sorella}}]{stripe-vmc2}%
  \BibitemOpen
  \bibfield  {author} {\bibinfo {author} {\bibfnamefont {L.~F.}\ \bibnamefont
  {Tocchio}}, \bibinfo {author} {\bibfnamefont {F.}~\bibnamefont {Becca}},\
  and\ \bibinfo {author} {\bibfnamefont {S.}~\bibnamefont {Sorella}},\
  }\bibfield  {title} {\bibinfo {title} {Hidden {Mott} transition and large-$u$
  superconductivity in the two-dimensional {Hubbard} model},\ }\href
  {https://doi.org/10.1103/PhysRevB.94.195126} {\bibfield  {journal} {\bibinfo
  {journal} {Phys. Rev. B}\ }\textbf {\bibinfo {volume} {94}},\ \bibinfo
  {pages} {195126} (\bibinfo {year} {2016})}\BibitemShut {NoStop}%
\bibitem [{\citenamefont {Sorella}(2021)}]{stripe-afqmc}%
  \BibitemOpen
  \bibfield  {author} {\bibinfo {author} {\bibfnamefont {S.}~\bibnamefont
  {Sorella}},\ }\href@noop {} {\bibinfo {title} {The phase diagram of the
  {Hubbard} model by variational auxiliary field quantum monte carlo}}
  (\bibinfo {year} {2021}),\ \Eprint {https://arxiv.org/abs/2101.07045}
  {arXiv:2101.07045 [cond-mat.str-el]} \BibitemShut {NoStop}%
\bibitem [{\citenamefont {Chou}\ and\ \citenamefont {Lee}(2010)}]{vmctjstripe}%
  \BibitemOpen
  \bibfield  {author} {\bibinfo {author} {\bibfnamefont {C.-P.}\ \bibnamefont
  {Chou}}\ and\ \bibinfo {author} {\bibfnamefont {T.-K.}\ \bibnamefont {Lee}},\
  }\bibfield  {title} {\bibinfo {title} {Mechanism of formation of half-doped
  stripes in underdoped cuprates},\ }\href
  {https://doi.org/10.1103/PhysRevB.81.060503} {\bibfield  {journal} {\bibinfo
  {journal} {Phys. Rev. B}\ }\textbf {\bibinfo {volume} {81}},\ \bibinfo
  {pages} {060503(R)} (\bibinfo {year} {2010})}\BibitemShut {NoStop}%
\bibitem [{\citenamefont {Ido}\ \emph {et~al.}(2018{\natexlab{b}})\citenamefont
  {Ido}, \citenamefont {Ohgoe},\ and\ \citenamefont {Imada}}]{hubstp-Ido}%
  \BibitemOpen
  \bibfield  {author} {\bibinfo {author} {\bibfnamefont {K.}~\bibnamefont
  {Ido}}, \bibinfo {author} {\bibfnamefont {T.}~\bibnamefont {Ohgoe}},\ and\
  \bibinfo {author} {\bibfnamefont {M.}~\bibnamefont {Imada}},\ }\bibfield
  {title} {\bibinfo {title} {Competition among various charge-inhomogeneous
  states and $d$-wave superconducting state in {Hubbard} models on square
  lattices},\ }\href {https://doi.org/10.1103/PhysRevB.97.045138} {\bibfield
  {journal} {\bibinfo  {journal} {Phys. Rev. B}\ }\textbf {\bibinfo {volume}
  {97}},\ \bibinfo {pages} {045138} (\bibinfo {year}
  {2018}{\natexlab{b}})}\BibitemShut {NoStop}%
\bibitem [{\citenamefont {Tocchio}\ \emph {et~al.}(2019)\citenamefont
  {Tocchio}, \citenamefont {Montorsi},\ and\ \citenamefont
  {Becca}}]{hubstp-tocchio}%
  \BibitemOpen
  \bibfield  {author} {\bibinfo {author} {\bibfnamefont {L.~F.}\ \bibnamefont
  {Tocchio}}, \bibinfo {author} {\bibfnamefont {A.}~\bibnamefont {Montorsi}},\
  and\ \bibinfo {author} {\bibfnamefont {F.}~\bibnamefont {Becca}},\ }\bibfield
   {title} {\bibinfo {title} {{Metallic and insulating stripes and their
  relation with superconductivity in the doped {Hubbard} model}},\ }\href
  {https://doi.org/10.21468/SciPostPhys.7.2.021} {\bibfield  {journal}
  {\bibinfo  {journal} {SciPost Phys.}\ }\textbf {\bibinfo {volume} {7}},\
  \bibinfo {pages} {21} (\bibinfo {year} {2019})}\BibitemShut {NoStop}%
\bibitem [{\citenamefont {S\'en\'echal}\ \emph {et~al.}(2005)\citenamefont
  {S\'en\'echal}, \citenamefont {Lavertu}, \citenamefont {Marois},\ and\
  \citenamefont {Tremblay}}]{senechal2005}%
  \BibitemOpen
  \bibfield  {author} {\bibinfo {author} {\bibfnamefont {D.}~\bibnamefont
  {S\'en\'echal}}, \bibinfo {author} {\bibfnamefont {P.-L.}\ \bibnamefont
  {Lavertu}}, \bibinfo {author} {\bibfnamefont {M.-A.}\ \bibnamefont
  {Marois}},\ and\ \bibinfo {author} {\bibfnamefont {A.-M.~S.}\ \bibnamefont
  {Tremblay}},\ }\bibfield  {title} {\bibinfo {title} {Competition between
  antiferromagnetism and superconductivity in high-${T}_{c}$ cuprates},\ }\href
  {https://doi.org/10.1103/PhysRevLett.94.156404} {\bibfield  {journal}
  {\bibinfo  {journal} {Phys. Rev. Lett.}\ }\textbf {\bibinfo {volume} {94}},\
  \bibinfo {pages} {156404} (\bibinfo {year} {2005})}\BibitemShut {NoStop}%
\bibitem [{\citenamefont {Machida}(1989)}]{machida1}%
  \BibitemOpen
  \bibfield  {author} {\bibinfo {author} {\bibfnamefont {K.}~\bibnamefont
  {Machida}},\ }\bibfield  {title} {\bibinfo {title} {Magnetism in
  {La$_2$CuO$_4$} based compounds},\ }\href@noop {} {\bibfield  {journal}
  {\bibinfo  {journal} {Physica C: Superconductivity}\ }\textbf {\bibinfo
  {volume} {158}},\ \bibinfo {pages} {192} (\bibinfo {year}
  {1989})}\BibitemShut {NoStop}%
\bibitem [{\citenamefont {Xu}\ and\ \citenamefont
  {Grover}(2020)}]{xu2020competing}%
  \BibitemOpen
  \bibfield  {author} {\bibinfo {author} {\bibfnamefont {X.~Y.}\ \bibnamefont
  {Xu}}\ and\ \bibinfo {author} {\bibfnamefont {T.}~\bibnamefont {Grover}},\
  }\href@noop {} {\bibinfo {title} {Competing nodal d-wave superconductivity
  and antiferromagnetism: a quantum monte carlo study}} (\bibinfo {year}
  {2020}),\ \Eprint {https://arxiv.org/abs/2009.06644} {arXiv:2009.06644
  [cond-mat.str-el]} \BibitemShut {NoStop}%
\bibitem [{\citenamefont {Himeda}\ \emph {et~al.}(2002)\citenamefont {Himeda},
  \citenamefont {Kato},\ and\ \citenamefont {Ogata}}]{himeda2002stripe}%
  \BibitemOpen
  \bibfield  {author} {\bibinfo {author} {\bibfnamefont {A.}~\bibnamefont
  {Himeda}}, \bibinfo {author} {\bibfnamefont {T.}~\bibnamefont {Kato}},\ and\
  \bibinfo {author} {\bibfnamefont {M.}~\bibnamefont {Ogata}},\ }\bibfield
  {title} {\bibinfo {title} {Stripe states with spatially oscillating d-wave
  superconductivity in the two-dimensional
  $\mathit{t}\ensuremath{-}\mathit{t'}\ensuremath{-}\mathit{J}$ model},\
  }\href@noop {} {\bibfield  {journal} {\bibinfo  {journal} {Physical review
  letters}\ }\textbf {\bibinfo {volume} {88}},\ \bibinfo {pages} {117001}
  (\bibinfo {year} {2002})}\BibitemShut {NoStop}%
\bibitem [{\citenamefont {Jiang}\ and\ \citenamefont
  {Kivelson}(2022)}]{jiang2022stripe}%
  \BibitemOpen
  \bibfield  {author} {\bibinfo {author} {\bibfnamefont {H.-C.}\ \bibnamefont
  {Jiang}}\ and\ \bibinfo {author} {\bibfnamefont {S.~A.}\ \bibnamefont
  {Kivelson}},\ }\bibfield  {title} {\bibinfo {title} {Stripe order enhanced
  superconductivity in the hubbard model},\ }\href@noop {} {\bibfield
  {journal} {\bibinfo  {journal} {Proceedings of the National Academy of
  Sciences}\ }\textbf {\bibinfo {volume} {119}} (\bibinfo {year}
  {2022})}\BibitemShut {NoStop}%
\bibitem [{\citenamefont {Xu}\ \emph {et~al.}(2022)\citenamefont {Xu},
  \citenamefont {Shi}, \citenamefont {Vitali}, \citenamefont {Qin},\ and\
  \citenamefont {Zhang}}]{xu2022stripes}%
  \BibitemOpen
  \bibfield  {author} {\bibinfo {author} {\bibfnamefont {H.}~\bibnamefont
  {Xu}}, \bibinfo {author} {\bibfnamefont {H.}~\bibnamefont {Shi}}, \bibinfo
  {author} {\bibfnamefont {E.}~\bibnamefont {Vitali}}, \bibinfo {author}
  {\bibfnamefont {M.}~\bibnamefont {Qin}},\ and\ \bibinfo {author}
  {\bibfnamefont {S.}~\bibnamefont {Zhang}},\ }\bibfield  {title} {\bibinfo
  {title} {Stripes and spin-density waves in the doped two-dimensional hubbard
  model: Ground state phase diagram},\ }\href@noop {} {\bibfield  {journal}
  {\bibinfo  {journal} {Physical Review Research}\ }\textbf {\bibinfo {volume}
  {4}},\ \bibinfo {pages} {013239} (\bibinfo {year} {2022})}\BibitemShut
  {NoStop}%
\bibitem [{\citenamefont {Huang}\ \emph {et~al.}(2022)\citenamefont {Huang},
  \citenamefont {Liu}, \citenamefont {Wang}, \citenamefont {Jiang},
  \citenamefont {Mai}, \citenamefont {Maier}, \citenamefont {Johnston},
  \citenamefont {Moritz},\ and\ \citenamefont
  {Devereaux}}]{huang2022intertwined}%
  \BibitemOpen
  \bibfield  {author} {\bibinfo {author} {\bibfnamefont {E.~W.}\ \bibnamefont
  {Huang}}, \bibinfo {author} {\bibfnamefont {T.}~\bibnamefont {Liu}}, \bibinfo
  {author} {\bibfnamefont {W.~O.}\ \bibnamefont {Wang}}, \bibinfo {author}
  {\bibfnamefont {H.-C.}\ \bibnamefont {Jiang}}, \bibinfo {author}
  {\bibfnamefont {P.}~\bibnamefont {Mai}}, \bibinfo {author} {\bibfnamefont
  {T.~A.}\ \bibnamefont {Maier}}, \bibinfo {author} {\bibfnamefont
  {S.}~\bibnamefont {Johnston}}, \bibinfo {author} {\bibfnamefont
  {B.}~\bibnamefont {Moritz}},\ and\ \bibinfo {author} {\bibfnamefont {T.~P.}\
  \bibnamefont {Devereaux}},\ }\href
  {https://doi.org/10.48550/ARXIV.2202.08845} {\bibinfo {title} {Fluctuating
  intertwined stripes in the strange metal regime of the hubbard model}}
  (\bibinfo {year} {2022})\BibitemShut {NoStop}%
\bibitem [{\citenamefont {Mai}\ \emph {et~al.}(2022)\citenamefont {Mai},
  \citenamefont {Karakuzu}, \citenamefont {Balduzzi}, \citenamefont
  {Johnston},\ and\ \citenamefont {Maier}}]{mai2022intertwined}%
  \BibitemOpen
  \bibfield  {author} {\bibinfo {author} {\bibfnamefont {P.}~\bibnamefont
  {Mai}}, \bibinfo {author} {\bibfnamefont {S.}~\bibnamefont {Karakuzu}},
  \bibinfo {author} {\bibfnamefont {G.}~\bibnamefont {Balduzzi}}, \bibinfo
  {author} {\bibfnamefont {S.}~\bibnamefont {Johnston}},\ and\ \bibinfo
  {author} {\bibfnamefont {T.~A.}\ \bibnamefont {Maier}},\ }\bibfield  {title}
  {\bibinfo {title} {Intertwined spin, charge, and pair correlations in the
  two-dimensional hubbard model in the thermodynamic limit},\ }\href@noop {}
  {\bibfield  {journal} {\bibinfo  {journal} {Proceedings of the National
  Academy of Sciences}\ }\textbf {\bibinfo {volume} {119}},\ \bibinfo {pages}
  {e2112806119} (\bibinfo {year} {2022})}\BibitemShut {NoStop}%
\bibitem [{\citenamefont {LeBlanc}\ \emph {et~al.}(2015)\citenamefont
  {LeBlanc}, \citenamefont {Antipov}, \citenamefont {Becca}, \citenamefont
  {Bulik}, \citenamefont {Chan}, \citenamefont {Chung}, \citenamefont {Deng},
  \citenamefont {Ferrero}, \citenamefont {Henderson}, \citenamefont
  {Jim\'enez-Hoyos}, \citenamefont {Kozik}, \citenamefont {Liu}, \citenamefont
  {Millis}, \citenamefont {Prokof'ev}, \citenamefont {Qin}, \citenamefont
  {Scuseria}, \citenamefont {Shi}, \citenamefont {Svistunov}, \citenamefont
  {Tocchio}, \citenamefont {Tupitsyn}, \citenamefont {White}, \citenamefont
  {Zhang}, \citenamefont {Zheng}, \citenamefont {Zhu},\ and\ \citenamefont
  {Gull}}]{hub-bench2015}%
  \BibitemOpen
  \bibfield  {author} {\bibinfo {author} {\bibfnamefont {J.~P.~F.}\
  \bibnamefont {LeBlanc}}, \bibinfo {author} {\bibfnamefont {A.~E.}\
  \bibnamefont {Antipov}}, \bibinfo {author} {\bibfnamefont {F.}~\bibnamefont
  {Becca}}, \bibinfo {author} {\bibfnamefont {I.~W.}\ \bibnamefont {Bulik}},
  \bibinfo {author} {\bibfnamefont {G.~K.-L.}\ \bibnamefont {Chan}}, \bibinfo
  {author} {\bibfnamefont {C.-M.}\ \bibnamefont {Chung}}, \bibinfo {author}
  {\bibfnamefont {Y.}~\bibnamefont {Deng}}, \bibinfo {author} {\bibfnamefont
  {M.}~\bibnamefont {Ferrero}}, \bibinfo {author} {\bibfnamefont {T.~M.}\
  \bibnamefont {Henderson}}, \bibinfo {author} {\bibfnamefont {C.~A.}\
  \bibnamefont {Jim\'enez-Hoyos}}, \bibinfo {author} {\bibfnamefont
  {E.}~\bibnamefont {Kozik}}, \bibinfo {author} {\bibfnamefont {X.-W.}\
  \bibnamefont {Liu}}, \bibinfo {author} {\bibfnamefont {A.~J.}\ \bibnamefont
  {Millis}}, \bibinfo {author} {\bibfnamefont {N.~V.}\ \bibnamefont
  {Prokof'ev}}, \bibinfo {author} {\bibfnamefont {M.}~\bibnamefont {Qin}},
  \bibinfo {author} {\bibfnamefont {G.~E.}\ \bibnamefont {Scuseria}}, \bibinfo
  {author} {\bibfnamefont {H.}~\bibnamefont {Shi}}, \bibinfo {author}
  {\bibfnamefont {B.~V.}\ \bibnamefont {Svistunov}}, \bibinfo {author}
  {\bibfnamefont {L.~F.}\ \bibnamefont {Tocchio}}, \bibinfo {author}
  {\bibfnamefont {I.~S.}\ \bibnamefont {Tupitsyn}}, \bibinfo {author}
  {\bibfnamefont {S.~R.}\ \bibnamefont {White}}, \bibinfo {author}
  {\bibfnamefont {S.}~\bibnamefont {Zhang}}, \bibinfo {author} {\bibfnamefont
  {B.-X.}\ \bibnamefont {Zheng}}, \bibinfo {author} {\bibfnamefont
  {Z.}~\bibnamefont {Zhu}},\ and\ \bibinfo {author} {\bibfnamefont
  {E.}~\bibnamefont {Gull}} (\bibinfo {collaboration} {Simons Collaboration on
  the Many-Electron Problem}),\ }\bibfield  {title} {\bibinfo {title}
  {Solutions of the two-dimensional {Hubbard} model: Benchmarks and results
  from a wide range of numerical algorithms},\ }\href
  {https://doi.org/10.1103/PhysRevX.5.041041} {\bibfield  {journal} {\bibinfo
  {journal} {Phys. Rev. X}\ }\textbf {\bibinfo {volume} {5}},\ \bibinfo {pages}
  {041041} (\bibinfo {year} {2015})}\BibitemShut {NoStop}%
\bibitem [{\citenamefont {Zheng}\ \emph {et~al.}(2017)\citenamefont {Zheng},
  \citenamefont {Chung}, \citenamefont {Corboz}, \citenamefont {Ehlers},
  \citenamefont {Qin}, \citenamefont {Noack}, \citenamefont {Shi},
  \citenamefont {White}, \citenamefont {Zhang},\ and\ \citenamefont
  {Chan}}]{stripehubbard}%
  \BibitemOpen
  \bibfield  {author} {\bibinfo {author} {\bibfnamefont {B.-X.}\ \bibnamefont
  {Zheng}}, \bibinfo {author} {\bibfnamefont {C.-M.}\ \bibnamefont {Chung}},
  \bibinfo {author} {\bibfnamefont {P.}~\bibnamefont {Corboz}}, \bibinfo
  {author} {\bibfnamefont {G.}~\bibnamefont {Ehlers}}, \bibinfo {author}
  {\bibfnamefont {M.-P.}\ \bibnamefont {Qin}}, \bibinfo {author} {\bibfnamefont
  {R.~M.}\ \bibnamefont {Noack}}, \bibinfo {author} {\bibfnamefont
  {H.}~\bibnamefont {Shi}}, \bibinfo {author} {\bibfnamefont {S.~R.}\
  \bibnamefont {White}}, \bibinfo {author} {\bibfnamefont {S.}~\bibnamefont
  {Zhang}},\ and\ \bibinfo {author} {\bibfnamefont {G.~K.-L.}\ \bibnamefont
  {Chan}},\ }\bibfield  {title} {\bibinfo {title} {Stripe order in the
  underdoped region of the two-dimensional {Hubbard} model},\ }\href@noop {}
  {\bibfield  {journal} {\bibinfo  {journal} {Science}\ }\textbf {\bibinfo
  {volume} {358}},\ \bibinfo {pages} {1155} (\bibinfo {year}
  {2017})}\BibitemShut {NoStop}%
\bibitem [{\citenamefont {Jiang}\ \emph {et~al.}(2021)\citenamefont {Jiang},
  \citenamefont {Scalapino},\ and\ \citenamefont {White}}]{tt'j}%
  \BibitemOpen
  \bibfield  {author} {\bibinfo {author} {\bibfnamefont {S.}~\bibnamefont
  {Jiang}}, \bibinfo {author} {\bibfnamefont {D.~J.}\ \bibnamefont
  {Scalapino}},\ and\ \bibinfo {author} {\bibfnamefont {S.~R.}\ \bibnamefont
  {White}},\ }\bibfield  {title} {\bibinfo {title} {Ground-state phase diagram
  of the $t$-$t'$-${J}$ model},\ }\bibfield  {journal} {\bibinfo  {journal}
  {Proceedings of the National Academy of Sciences}\ }\textbf {\bibinfo
  {volume} {118}},\ \href {https://doi.org/10.1073/pnas.2109978118}
  {10.1073/pnas.2109978118} (\bibinfo {year} {2021})\BibitemShut {NoStop}%
\bibitem [{\citenamefont {Gong}\ \emph {et~al.}(2021)\citenamefont {Gong},
  \citenamefont {Zhu},\ and\ \citenamefont {Sheng}}]{sheng}%
  \BibitemOpen
  \bibfield  {author} {\bibinfo {author} {\bibfnamefont {S.}~\bibnamefont
  {Gong}}, \bibinfo {author} {\bibfnamefont {W.}~\bibnamefont {Zhu}},\ and\
  \bibinfo {author} {\bibfnamefont {D.~N.}\ \bibnamefont {Sheng}},\ }\bibfield
  {title} {\bibinfo {title} {Robust $d$-wave superconductivity in the
  square-lattice $t$-${J}$ model},\ }\href
  {https://doi.org/10.1103/PhysRevLett.127.097003} {\bibfield  {journal}
  {\bibinfo  {journal} {Phys. Rev. Lett.}\ }\textbf {\bibinfo {volume} {127}},\
  \bibinfo {pages} {097003} (\bibinfo {year} {2021})}\BibitemShut {NoStop}%
\bibitem [{\citenamefont {Greene}\ \emph {et~al.}(2020)\citenamefont {Greene},
  \citenamefont {Mandal}, \citenamefont {Poniatowski},\ and\ \citenamefont
  {Sarkar}}]{greene2020strange}%
  \BibitemOpen
  \bibfield  {author} {\bibinfo {author} {\bibfnamefont {R.~L.}\ \bibnamefont
  {Greene}}, \bibinfo {author} {\bibfnamefont {P.~R.}\ \bibnamefont {Mandal}},
  \bibinfo {author} {\bibfnamefont {N.~R.}\ \bibnamefont {Poniatowski}},\ and\
  \bibinfo {author} {\bibfnamefont {T.}~\bibnamefont {Sarkar}},\ }\bibfield
  {title} {\bibinfo {title} {The strange metal state of the electron-doped
  cuprates},\ }\href@noop {} {\bibfield  {journal} {\bibinfo  {journal} {Annual
  Review of Condensed Matter Physics}\ }\textbf {\bibinfo {volume} {11}},\
  \bibinfo {pages} {213} (\bibinfo {year} {2020})}\BibitemShut {NoStop}%
\bibitem [{\citenamefont {Lee}\ \emph {et~al.}(2006)\citenamefont {Lee},
  \citenamefont {Nagaosa},\ and\ \citenamefont {Wen}}]{lee2006doping}%
  \BibitemOpen
  \bibfield  {author} {\bibinfo {author} {\bibfnamefont {P.~A.}\ \bibnamefont
  {Lee}}, \bibinfo {author} {\bibfnamefont {N.}~\bibnamefont {Nagaosa}},\ and\
  \bibinfo {author} {\bibfnamefont {X.-G.}\ \bibnamefont {Wen}},\ }\bibfield
  {title} {\bibinfo {title} {Doping a mott insulator: Physics of
  high-temperature superconductivity},\ }\href@noop {} {\bibfield  {journal}
  {\bibinfo  {journal} {Reviews of modern physics}\ }\textbf {\bibinfo {volume}
  {78}},\ \bibinfo {pages} {17} (\bibinfo {year} {2006})}\BibitemShut {NoStop}%
\bibitem [{\citenamefont {Leung}\ \emph {et~al.}(1997)\citenamefont {Leung},
  \citenamefont {Wells},\ and\ \citenamefont {Gooding}}]{leung1997}%
  \BibitemOpen
  \bibfield  {author} {\bibinfo {author} {\bibfnamefont {P.~W.}\ \bibnamefont
  {Leung}}, \bibinfo {author} {\bibfnamefont {B.~O.}\ \bibnamefont {Wells}},\
  and\ \bibinfo {author} {\bibfnamefont {R.~J.}\ \bibnamefont {Gooding}},\
  }\bibfield  {title} {\bibinfo {title} {Comparison of 32-site
  exact-diagonalization results and arpes spectral functions for the
  antiferromagnetic insulator sr 2 cuo 2 cl 2},\ }\href@noop {} {\bibfield
  {journal} {\bibinfo  {journal} {Physical Review B}\ }\textbf {\bibinfo
  {volume} {56}},\ \bibinfo {pages} {6320} (\bibinfo {year}
  {1997})}\BibitemShut {NoStop}%
\bibitem [{\citenamefont {Xiang}\ and\ \citenamefont
  {Wheatley}(1996)}]{xiang1996}%
  \BibitemOpen
  \bibfield  {author} {\bibinfo {author} {\bibfnamefont {T.}~\bibnamefont
  {Xiang}}\ and\ \bibinfo {author} {\bibfnamefont {J.}~\bibnamefont
  {Wheatley}},\ }\bibfield  {title} {\bibinfo {title} {Quasiparticle energy
  dispersion in doped two-dimensional quantum antiferromagnets},\ }\href@noop
  {} {\bibfield  {journal} {\bibinfo  {journal} {Physical Review B}\ }\textbf
  {\bibinfo {volume} {54}},\ \bibinfo {pages} {R12653} (\bibinfo {year}
  {1996})}\BibitemShut {NoStop}%
\bibitem [{\citenamefont {Zhang}\ and\ \citenamefont {Rice}(1988)}]{zhangrice}%
  \BibitemOpen
  \bibfield  {author} {\bibinfo {author} {\bibfnamefont {F.}~\bibnamefont
  {Zhang}}\ and\ \bibinfo {author} {\bibfnamefont {T.}~\bibnamefont {Rice}},\
  }\bibfield  {title} {\bibinfo {title} {Effective hamiltonian for the
  superconducting cu oxides},\ }\href@noop {} {\bibfield  {journal} {\bibinfo
  {journal} {Physical Review B}\ }\textbf {\bibinfo {volume} {37}},\ \bibinfo
  {pages} {3759} (\bibinfo {year} {1988})}\BibitemShut {NoStop}%
\bibitem [{\citenamefont {Belinicher}\ \emph {et~al.}(1996)\citenamefont
  {Belinicher}, \citenamefont {Chernyshev},\ and\ \citenamefont
  {Shubin}}]{sasha-tt'J}%
  \BibitemOpen
  \bibfield  {author} {\bibinfo {author} {\bibfnamefont {V.~I.}\ \bibnamefont
  {Belinicher}}, \bibinfo {author} {\bibfnamefont {A.~L.}\ \bibnamefont
  {Chernyshev}},\ and\ \bibinfo {author} {\bibfnamefont {V.~A.}\ \bibnamefont
  {Shubin}},\ }\bibfield  {title} {\bibinfo {title} {Generalized
  t-t\ensuremath{'}-j model: Parameters and single-particle spectrum for
  electrons and holes in copper oxides},\ }\href
  {https://doi.org/10.1103/PhysRevB.53.335} {\bibfield  {journal} {\bibinfo
  {journal} {Phys. Rev. B}\ }\textbf {\bibinfo {volume} {53}},\ \bibinfo
  {pages} {335} (\bibinfo {year} {1996})}\BibitemShut {NoStop}%
\bibitem [{\citenamefont {Fishman}\ \emph {et~al.}(2020)\citenamefont
  {Fishman}, \citenamefont {White},\ and\ \citenamefont
  {Stoudenmire}}]{itensor}%
  \BibitemOpen
  \bibfield  {author} {\bibinfo {author} {\bibfnamefont {M.}~\bibnamefont
  {Fishman}}, \bibinfo {author} {\bibfnamefont {S.~R.}\ \bibnamefont {White}},\
  and\ \bibinfo {author} {\bibfnamefont {E.~M.}\ \bibnamefont {Stoudenmire}},\
  }\href@noop {} {\bibinfo {title} {The \mbox{ITensor} software library for
  tensor network calculations}} (\bibinfo {year} {2020}),\ \Eprint
  {https://arxiv.org/abs/2007.14822} {arXiv:2007.14822} \BibitemShut {NoStop}%
\bibitem [{sm()}]{sm}%
  \BibitemOpen
  \href@noop {} {}\bibinfo {note} {See Supplemental Material at [] for the rest
  of scan calculations.}\BibitemShut {Stop}%
\bibitem [{\citenamefont {Jiang}\ and\ \citenamefont
  {Kivelson}(2021)}]{jiang2021high}%
  \BibitemOpen
  \bibfield  {author} {\bibinfo {author} {\bibfnamefont {H.-C.}\ \bibnamefont
  {Jiang}}\ and\ \bibinfo {author} {\bibfnamefont {S.~A.}\ \bibnamefont
  {Kivelson}},\ }\bibfield  {title} {\bibinfo {title} {High temperature
  superconductivity in a lightly doped quantum spin liquid},\ }\href@noop {}
  {\bibfield  {journal} {\bibinfo  {journal} {Physical Review Letters}\
  }\textbf {\bibinfo {volume} {127}},\ \bibinfo {pages} {097002} (\bibinfo
  {year} {2021})}\BibitemShut {NoStop}%
\bibitem [{\citenamefont {White}\ and\ \citenamefont
  {Scalapino}(2015)}]{white2015doping}%
  \BibitemOpen
  \bibfield  {author} {\bibinfo {author} {\bibfnamefont {S.~R.}\ \bibnamefont
  {White}}\ and\ \bibinfo {author} {\bibfnamefont {D.}~\bibnamefont
  {Scalapino}},\ }\bibfield  {title} {\bibinfo {title} {Doping asymmetry and
  striping in a three-orbital cuo 2 hubbard model},\ }\href@noop {} {\bibfield
  {journal} {\bibinfo  {journal} {Physical Review B}\ }\textbf {\bibinfo
  {volume} {92}},\ \bibinfo {pages} {205112} (\bibinfo {year}
  {2015})}\BibitemShut {NoStop}%
\bibitem [{\citenamefont {Cui}\ \emph {et~al.}(2020)\citenamefont {Cui},
  \citenamefont {Sun}, \citenamefont {Ray}, \citenamefont {Zheng},
  \citenamefont {Sun},\ and\ \citenamefont {Chan}}]{cui2020}%
  \BibitemOpen
  \bibfield  {author} {\bibinfo {author} {\bibfnamefont {Z.-H.}\ \bibnamefont
  {Cui}}, \bibinfo {author} {\bibfnamefont {C.}~\bibnamefont {Sun}}, \bibinfo
  {author} {\bibfnamefont {U.}~\bibnamefont {Ray}}, \bibinfo {author}
  {\bibfnamefont {B.-X.}\ \bibnamefont {Zheng}}, \bibinfo {author}
  {\bibfnamefont {Q.}~\bibnamefont {Sun}},\ and\ \bibinfo {author}
  {\bibfnamefont {G.~K.-L.}\ \bibnamefont {Chan}},\ }\bibfield  {title}
  {\bibinfo {title} {Ground-state phase diagram of the three-band hubbard model
  from density matrix embedding theory},\ }\href@noop {} {\bibfield  {journal}
  {\bibinfo  {journal} {Physical Review Research}\ }\textbf {\bibinfo {volume}
  {2}},\ \bibinfo {pages} {043259} (\bibinfo {year} {2020})}\BibitemShut
  {NoStop}%
\bibitem [{\citenamefont {Mai}\ \emph {et~al.}(2021)\citenamefont {Mai},
  \citenamefont {Balduzzi}, \citenamefont {Johnston},\ and\ \citenamefont
  {Maier}}]{mai2021}%
  \BibitemOpen
  \bibfield  {author} {\bibinfo {author} {\bibfnamefont {P.}~\bibnamefont
  {Mai}}, \bibinfo {author} {\bibfnamefont {G.}~\bibnamefont {Balduzzi}},
  \bibinfo {author} {\bibfnamefont {S.}~\bibnamefont {Johnston}},\ and\
  \bibinfo {author} {\bibfnamefont {T.~A.}\ \bibnamefont {Maier}},\ }\bibfield
  {title} {\bibinfo {title} {Orbital structure of the effective pairing
  interaction in the high-temperature superconducting cuprates},\ }\href@noop
  {} {\bibfield  {journal} {\bibinfo  {journal} {npj Quantum Materials}\
  }\textbf {\bibinfo {volume} {6}},\ \bibinfo {pages} {1} (\bibinfo {year}
  {2021})}\BibitemShut {NoStop}%
\bibitem [{\citenamefont {Arrigoni}\ \emph {et~al.}(2009)\citenamefont
  {Arrigoni}, \citenamefont {Aichhorn}, \citenamefont {Daghofer},\ and\
  \citenamefont {Hanke}}]{arrigoni2009phase}%
  \BibitemOpen
  \bibfield  {author} {\bibinfo {author} {\bibfnamefont {E.}~\bibnamefont
  {Arrigoni}}, \bibinfo {author} {\bibfnamefont {M.}~\bibnamefont {Aichhorn}},
  \bibinfo {author} {\bibfnamefont {M.}~\bibnamefont {Daghofer}},\ and\
  \bibinfo {author} {\bibfnamefont {W.}~\bibnamefont {Hanke}},\ }\bibfield
  {title} {\bibinfo {title} {Phase diagram and single-particle spectrum of cuo2
  high-tc layers: variational cluster approach to the three-band hubbard
  model},\ }\href@noop {} {\bibfield  {journal} {\bibinfo  {journal} {New
  Journal of Physics}\ }\textbf {\bibinfo {volume} {11}},\ \bibinfo {pages}
  {055066} (\bibinfo {year} {2009})}\BibitemShut {NoStop}%
\bibitem [{\citenamefont {Brookes}\ \emph {et~al.}(2001)\citenamefont
  {Brookes}, \citenamefont {Ghiringhelli}, \citenamefont {Tjernberg},
  \citenamefont {Tjeng}, \citenamefont {Mizokawa}, \citenamefont {Li},\ and\
  \citenamefont {Menovsky}}]{brookes2001detection}%
  \BibitemOpen
  \bibfield  {author} {\bibinfo {author} {\bibfnamefont {N.}~\bibnamefont
  {Brookes}}, \bibinfo {author} {\bibfnamefont {G.}~\bibnamefont
  {Ghiringhelli}}, \bibinfo {author} {\bibfnamefont {O.}~\bibnamefont
  {Tjernberg}}, \bibinfo {author} {\bibfnamefont {L.}~\bibnamefont {Tjeng}},
  \bibinfo {author} {\bibfnamefont {T.}~\bibnamefont {Mizokawa}}, \bibinfo
  {author} {\bibfnamefont {T.}~\bibnamefont {Li}},\ and\ \bibinfo {author}
  {\bibfnamefont {A.}~\bibnamefont {Menovsky}},\ }\bibfield  {title} {\bibinfo
  {title} {Detection of zhang-rice singlets using spin-polarized
  photoemission},\ }\href@noop {} {\bibfield  {journal} {\bibinfo  {journal}
  {Physical Review Letters}\ }\textbf {\bibinfo {volume} {87}},\ \bibinfo
  {pages} {237003} (\bibinfo {year} {2001})}\BibitemShut {NoStop}%
\bibitem [{\citenamefont {Tjeng}\ \emph {et~al.}(1997)\citenamefont {Tjeng},
  \citenamefont {Sinkovic}, \citenamefont {Brookes}, \citenamefont {Goedkoop},
  \citenamefont {Hesper}, \citenamefont {Pellegrin}, \citenamefont {De~Groot},
  \citenamefont {Altieri}, \citenamefont {Hulbert}, \citenamefont {Shekel}
  \emph {et~al.}}]{tjeng1997spin}%
  \BibitemOpen
  \bibfield  {author} {\bibinfo {author} {\bibfnamefont {L.}~\bibnamefont
  {Tjeng}}, \bibinfo {author} {\bibfnamefont {B.}~\bibnamefont {Sinkovic}},
  \bibinfo {author} {\bibfnamefont {N.}~\bibnamefont {Brookes}}, \bibinfo
  {author} {\bibfnamefont {J.}~\bibnamefont {Goedkoop}}, \bibinfo {author}
  {\bibfnamefont {R.}~\bibnamefont {Hesper}}, \bibinfo {author} {\bibfnamefont
  {E.}~\bibnamefont {Pellegrin}}, \bibinfo {author} {\bibfnamefont
  {F.}~\bibnamefont {De~Groot}}, \bibinfo {author} {\bibfnamefont
  {S.}~\bibnamefont {Altieri}}, \bibinfo {author} {\bibfnamefont
  {S.}~\bibnamefont {Hulbert}}, \bibinfo {author} {\bibfnamefont
  {E.}~\bibnamefont {Shekel}}, \emph {et~al.},\ }\bibfield  {title} {\bibinfo
  {title} {Spin-resolved photoemission on anti-ferromagnets: direct observation
  of zhang-rice singlets in cuo},\ }\href@noop {} {\bibfield  {journal}
  {\bibinfo  {journal} {Physical review letters}\ }\textbf {\bibinfo {volume}
  {78}},\ \bibinfo {pages} {1126} (\bibinfo {year} {1997})}\BibitemShut
  {NoStop}%
\bibitem [{\citenamefont {Harada}\ \emph {et~al.}(2002)\citenamefont {Harada},
  \citenamefont {Okada}, \citenamefont {Eguchi}, \citenamefont {Kotani},
  \citenamefont {Takagi}, \citenamefont {Takeuchi},\ and\ \citenamefont
  {Shin}}]{harada2002unique}%
  \BibitemOpen
  \bibfield  {author} {\bibinfo {author} {\bibfnamefont {Y.}~\bibnamefont
  {Harada}}, \bibinfo {author} {\bibfnamefont {K.}~\bibnamefont {Okada}},
  \bibinfo {author} {\bibfnamefont {R.}~\bibnamefont {Eguchi}}, \bibinfo
  {author} {\bibfnamefont {A.}~\bibnamefont {Kotani}}, \bibinfo {author}
  {\bibfnamefont {H.}~\bibnamefont {Takagi}}, \bibinfo {author} {\bibfnamefont
  {T.}~\bibnamefont {Takeuchi}},\ and\ \bibinfo {author} {\bibfnamefont
  {S.}~\bibnamefont {Shin}},\ }\bibfield  {title} {\bibinfo {title} {Unique
  identification of zhang-rice singlet excitation in sr 2 cuo 2 cl 2 mediated
  by the o 1 s core hole: Symmetry-selective resonant soft x-ray raman
  scattering study},\ }\href@noop {} {\bibfield  {journal} {\bibinfo  {journal}
  {Physical Review B}\ }\textbf {\bibinfo {volume} {66}},\ \bibinfo {pages}
  {165104} (\bibinfo {year} {2002})}\BibitemShut {NoStop}%
\end{thebibliography}%

\end{document}